\documentclass[aps,prl,twocolumn,superscriptaddress,showpacs]{revtex4}
\usepackage[dvips]{graphicx}
\usepackage{dcolumn}
\usepackage{bm}
\begin{document}
%\draft
\preprint{}
\title{
An orbital glass state of a nearly metallic spinel, CoV$_2$O$_4$
} 
\author{R. Koborinai}
\affiliation{Department of Physics, Waseda University, Tokyo 169-8555, Japan}
\author{S.E. Dissanayake}
\thanks{Email: sachithd83@gmail.com}
\affiliation{Department of Physics, University of Virginia, Charlottesville, Virginia 
22904, USA}
\author{M. Reehuis}
\affiliation{Helmholtz-Zentrum f\"ur Materialien und Energie, 14109 Berlin, 
Germany}
\author{M. Matsuda}
\affiliation{Quantum Condensed Matter Division, Oak Ridge National Laboratory, Oak 
Ridge, TN 37831, U.S.A}
\author{S. -H. Lee}
\affiliation{Department of Physics, University of Virginia, Charlottesville, Virginia 
22904, USA}
\author{T. Katsufuji}
\thanks{Email: katsuf@waseda.jp}
\affiliation{Department of Physics, Waseda University, Tokyo 169-8555, Japan}
\affiliation{Kagami Memorial Research Institute for Materials Science and Technology, 
Waseda University, Tokyo 169-0051, Japan} 

\date{\today}

\begin{abstract}
Strain, magnetization, and unpolarized and polarized neutron diffraction measurements were performed to study 
the magnetic and structural properties of spinel CoV$_{2}$O$_{4}$. Magnetostriction 
measurements indicate that a subtle distortion of the crystal along the direction of 
magnetization, $\Delta L/L \sim 10^{-4}$, exists and varies from elongation to contraction in a 
second order fashion upon cooling. Unpolarized and polarized single-crystal neutron 
experiments indicate that upon cooling the ferrimagnetic structure changes from collinear to 
noncollinear at $T \sim 90$ K, where the elongation of the crystal is maximized. These 
results imply the existence of an orbital glassy state in the nearly metallic frustrated magnet CoV$_{2}$O$_{4}$.
\end{abstract}

\maketitle
The interplay between magnetic frustration and orbital degree of freedom has been 
extensively studied in various transition-metal oxides. Among them, spinel vanadates 
$A$V$_{2}$O$_{4}$, in which the magnetic V$^{3+}$ [$t_{2g}^{2}$] ions form a highly frustrated three-dimensional corner-sharing network of tetrahedra, are model systems to search for 
novel emergent phases. With non-magnetic $A^{2+}$ ions 
($A$ = Zn, Cd), upon cooling the system undergoes a first order structural phase transition 
due to an orbital long range order followed by a bi-partite magnetic order\cite{ueda97,lee04,zhang06}. With magnetic $A^{2+}$ ions such as Mn,Fe 
and Co, additional spin exchange interactions between $A^{2+}$ and V$^{3+}$ ions come into play to exhibit more complex behaviors. 
For example, the insulating MnV$_{2}$O$_{4}$ 
\cite{Suzuki07,Zhou07,Chung08,Garlea08,Hardy08,Baek09,Sarkar09,Chern10,Nii13,Gleason14} 
exhibits a first order structural phase transition from 
cubic to tetragonal (with a shorter $c$ axis) and ferrimagnetic order with a 
noncollinear V spin structure \cite{Garlea08} simultaneously at $T_{\rm C}=57$ K. 
This phase transition is dominated by the Kugel-Khomskii-type interactions that are the 
intersite interactions between orbital and spin degrees of freedom of the V ions.  
Another insulating FeV$_{2}$O$_{4}$ \cite{Katsufuji08,Sarkar11,MacDougall12,Zhang12,Nii12,Kang12,Kawaguchi13,Huang14,Choudhury14} exhibits, upon cooling, 
successive structural phase transitions from a cubic to a high-temperature 
tetragonal to an orthorhombic, and to a low-temperature tetragonal phase (with a longer 
$c$ axis), which are caused not only by the orbital degree of freedom of V ions but also of Fe$^{2+}$ (3d$^6$; $e_g^3t_{2g}^{3}$) ions \cite{Katsufuji08}.

The spinel CoV$_{2}$O$_{4}$ [Co$^{2+}$: $e_{g}^{4}t_{2g}^{3}$, V$^{3+}$: $t_{2g}^{2}$] 
provides a unique situation due to its close proximity to itineracy 
\cite{Kismarahardja11,Kiswandhi11,Huang12,Kaur14,Kismarahardja13,Ma14}. Unlike in the insulating AV$_2$O$_4$ (A = Mn, Fe), no observed crystal distortion has been observed in CoV$_2$O$_4$ down to 10 K by x-ray diffraction, while a similar ferrimagnetic order as in the insulating compounds was detected below $T_{\rm C} \simeq 150$ K by DC magnetization \cite{Kismarahardja11}. More recently, a neutron scattering study on Mn$_{1-x}$Co$_x$V$_{2}$O$_{4}$ proposed, by extrapolating from $x \leq 0.8$, the disappearance of orbital order for higher Co concentration ($x\geq0.8$) due to the enhancement of itineracy\cite{Ma14}. On the other hand, bulk magnetization ($M_{bulk} (T)$) data obtained from polycrystalline samples of CoV$_2$O$_4$ showed two cusps centered at 60 and 100 K, and specific heat, $C_V$, data exhibited one peak at 60 K and a broad peak centered at $T_{\rm C}$, which was attributed to 
a short-range orbital order \cite{Huang12}. This contradicts a previous single crystal study that reported one cusp in $M_{bulk} (T)$ at 75 K and no corresponding anomaly in $C_V$ \cite{Kiswandhi11}. More recently, dielectric measurements on a single crystal reported a contraction along the direction of the applied magnetic field below 30 K \cite{Kismarahardja13}.  Despite the conflicting results, the anomalies below $T_{\rm C}$ observed in the different measurements suggest that the orbital degree of freedom might be 
playing an important role also in this compound although the associated lattice 
distortion might be too small to be easily detected. This calls for the use of experimental probes 
that are more sensitive to such subtle changes in magnetic and structural properties 
that might exist in this nearly metallic spinel.

In this paper, we report on the strain, magnetization, and unpolarized and polarized neutron 
diffraction measurements of single crystals and polycrystalline samples of 
CoV$_{2}$O$_{4}$. The most salient feature of our data is that, upon cooling, the system 
undergoes weak lattice elongation below $T_{\rm C}$, of an order of $\Delta L_{\rm max}/L 
\sim 10^{-4}$ in a {\it second} order fashion, which differs from the strong, first 
order crystal distortions due to orbital long range order found in the other insulating vanadium 
spinels. Upon further cooling, the distortion continuously changes from elongation 
($\Delta L/L>0$) to contraction ($\Delta L/L<0$). The structural change is 
accompanied by changes in the ordered magnetic state. We argue that the unusual 
structural and magnetic behaviors are due to the system's proximity to the itineracy 
that incompletely suppresses the orbital degree of freedom leading to a glassy orbital state.

The single crystals of CoV$_{2}$O$_{4}$ were grown by the floating zone technique. When a 
polycrystalline rod with a stoichiometric amount of Co and V ($=1:2$) is used, a large single 
crystal of CoV$_{2}$O$_{4}$ cannot be grown because of the precipitation of the V$_{2}$O$_{3}$ impurity phase. Thus, the single crystals in this study were grown with extra Co. The Co:V ratios of the \#1 and \#2 single crystals were estimated by induction-coupled plasma 
analysis and found it to be 1.21:1.79 and 1.3:1.7, respectively. At the same time, polycrystalline samples with excess amounts of Co, 
Co$_{1+x}$V$_{2-x}$O$_{4}$ ($x=0,0.1$ and $0.2$) were synthesized for comparison in 
sealed quartz tubes. The bulk magnetization measurement on the samples was 
performed using a SQUID magnetometer. Strain measurements were performed using a 
strain-gauge technique. Neutron powder diffraction measurements were carried out on 
the BT1 powder diffractometer at the NIST Center for Neutron Research with a Cu(311) monochromator 
(${\lambda}$ = 1.5398 {\AA}), and the Rietveld refinements were carried out using the 
FULLPROF program. Single crystal neutron diffraction measurements were performed 
on the four-Circle Diffractometer E5 at the BERII reactor of the Helmholtz-Zentrum Berlin with neutron wavelengths of ${\lambda}$ = 2.4 {\AA} and 0.9 {\AA} . Polarized elastic neutron scattering experiments were 
performed at HB1 Polarized Triple-Axis Spectrometer at the High Flux Isotope Reactor, Oak Ridge National Laboratory with neutron 
energy of 13.5 meV. A vertical guide field of 3 T was applied using 8 T Vertical 
Asymmetric Field Cryomagnet.

\begin{figure}
\includegraphics[scale=0.47]{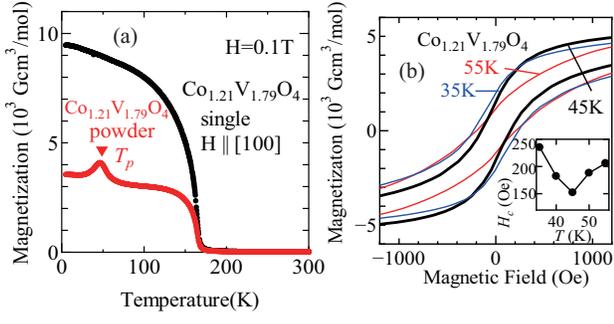}
\caption{
(Color online) (a) Temperature dependence of magnetization, $M_{bulk}(T)$ for a single crystal of 
CoV$_{2}$O$_{4}$ (\#1) with an applied field of 0.1 T along the [100] axis and that for 
powder sample obtained by grinding the single crystal. (b) Magnetization vs magnetic 
field for a ground single crystal of CoV$_{2}$O$_{4}$ (\#1) around $T_{p}=45$ K. The 
inset shows the temperature dependence of coercive field, $H_{c}$.
}
\end{figure}

Fig 1 (a) shows the temperature ($T$) dependence of $M_{bulk}$ obtained from a single crystal 
of CoV$_{2}$O$_{4}$ (\#1) with an applied magnetic field ($H$) of 0.1 T along the [100] 
direction. $M_{bulk}(T)$ increases below the ferrimagnetic transition temperature $T_{\rm C}$ $\sim 165$ K (black circles). Unlike in the case of polycrystalline samples, anomalies in $M_{bulk}(T)$ below $T_{\rm C}$ are barely visible in the single crystal data. When the same 
crystal was ground to a powder, however, $M_{bulk}(T)$ exhibits a clear anomaly at $T_{p}$ = 45 K (red circles). \cite{note0} The weak anomaly at $T_p$ can also be seen in $M_{bulk}(H)$ measured at several different temperatures around $T_p$. As shown in Fig. 1 (b), $M_{bulk}(H)$ exhibits hysteresis due to the presence of ferromagnetic components. The inset shows the 
$T$-dependence of the coercive field $H_{c}$ at which $M$ becomes zero. $H_{c} (T)$ exhibits a dip centered at $T_{p}=45$ K.

\begin{figure}
\includegraphics[scale=0.42]{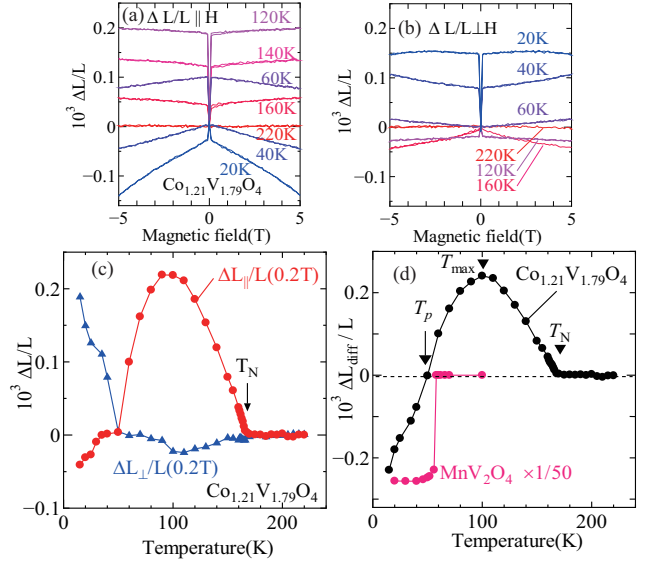}
\caption{
(Color online) Magnetostriction data obtained from the single crystal of CoV$_{2}$O$_{4}$ (\#1). $H$-dependence of strain ($\Delta L/L$) at various temperatures, (a) parallel to $H$, $\Delta L_{\parallel} /L$, and (b) perpendicular to $H$, $\Delta L_{\perp}/L$. (c) $T$-dependence of $\Delta L/L \parallel H$ (circles) and $\Delta L/L \perp 
H$ (triangles). (d) Difference in the strain, $\Delta L_{\rm diff}/L= (\Delta L_{\parallel} -\Delta L_{\perp})/L$  as a function of temperature. For comparison, 
$\textit{c/a}-1$ obtained by the x-ray diffraction measurement for MnV$_{2}$O$_{4}$ is 
also shown.
}
\end{figure}

Fig. 2 shows the strain ($\Delta L/L$) data obtained from the \#1 single crystal when the magnetic field is applied along the [100] direction, $H \parallel$ [100], as a function of $H$ and $T$. As shown in Fig. 2 (a) and (b), upon ramping up, the strain, both parallel ($\Delta L_{\parallel}/L$) and perpendicular ($\Delta L_{\perp}/L$) to $H$, do not show any response to $H$ for $T > T_{\rm C}$. Below $T_{\rm C}$, however, the strain exhibits a strong response. Interestingly, the strain response to $H$ shows an opposite behavior between the two temperature regimes, above and below $\sim$ 40 K. For 40 K $\lesssim T < T_{\rm C}$, as $|H|$ increases up to 0.2 T, $\Delta L_{\parallel}/L$ sharply increases while $\Delta L_{\perp}/L$ sharply decreases by a much smaller amount. Upon further ramping, the change in $\Delta L/L$ becomes gradual. For $T \lesssim$ 40 K, however, the strain response is opposite; upon ramping up to 0.2 T, $\Delta L_{\parallel}/L$ sharply decreases by a small amount while $\Delta L_{\perp}/L$ sharply increases by a much larger amount.

The $T$-dependence of the sharp response at low field, measured with $|H|=0.2$ T is shown in Fig. 2 (c). Upon cooling, 
$\Delta L_{\parallel}/L$ ($T,|H|=0.2$ T) (red circles) starts increasing gradually below $T_{\rm C}$, reaches its maximum value of $\sim 2 \times 10^{-4}$ at $\sim$ 100 K. Upon further cooling, it decreases, and becomes negative below $\sim$ 45 K to reach $- 4 \times 10^{-5}$ at $\sim$ 10 K. On the other hand, upon cooling, $\Delta L_{\perp}/L (T,|H|=0.2$ T) (blue triangles) decreases gradually to $\sim -2.5 \times 10^{-5}$ at $\sim 100$ K, then starts to increase to become positive below $\sim 45$ K and reaches its maximum value of $\sim 2 \times 10^{-4}$ at $\sim$ 10 K. The opposite strain response to $H$ between the two $T$ regimes, below and above $\sim 45$ K, is clearly illustrated in the $T-$dependence of the difference $\Delta L_{diff}/L = (\Delta L_{\parallel}-\Delta L_{\perp})/L$ with $|H|=0.2$ T shown as the black circles in Fig. 2 (d).

$\Delta L_{\rm diff}/L$ is the measure of distortion from cubic phase, 
$\frac{c}{a}-1$, where $a$ and $c$ are the lattice constants, assuming that the crystal 
is tetragonal and the magnetic moments are along the $c$ direction. It should be noted that the temperature, $\sim 45$ K, at which $\Delta L_{\rm diff}/L$ changes its sign coincides with $T_p$ where $M_{bulk}(T)$ exhibits a weak anomaly and $H_c (T)$ a minimum (see Fig. 1). We emphasize that the crystal distortion $\Delta L_{\rm diff}/L = \frac{c}{a}-1$ changes upon cooling from elongation to contraction, both of which occur in a second order transition. The second order crystal distortion in the nearly metallic CoV$_2$O$_4$ is in stark contrast with the sharp first-order contraction observed in the insulating MnV$_2$O$_4$ (red circles in Fig. 2 (d)).

\begin{figure}
\includegraphics[scale=0.28]{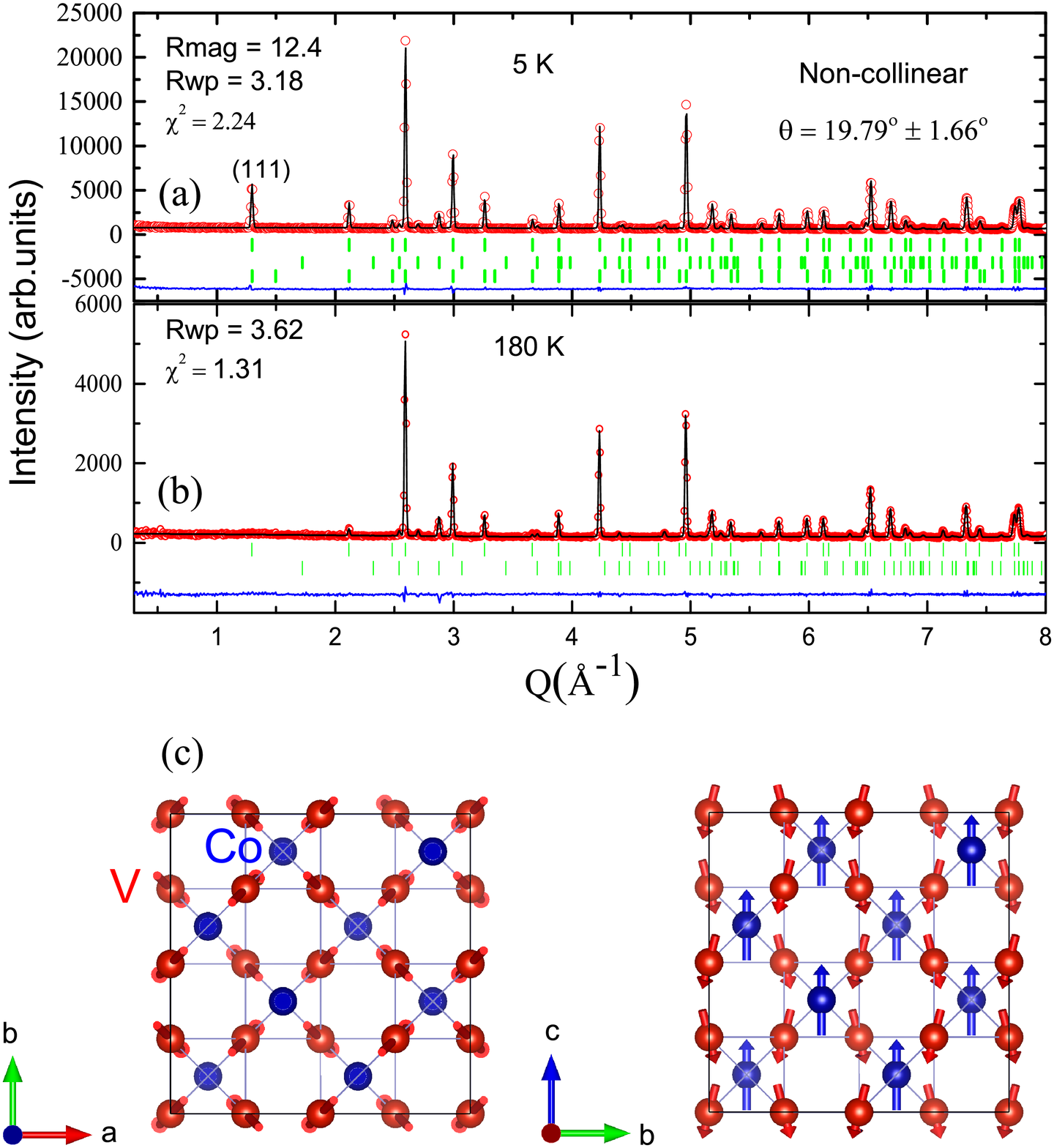}
\caption{(Color online) 
Neutron powder diffraction data of CoV$_2$O$_4$ measured at (a) 5 K and (b) 180 K. Circles are the 
experimental data and black lines represent the calculated intensities. Green bars 
represent nuclear and magnetic Bragg peak positions and blue lines indicate difference 
between experimental data and calculation. (c) A sketch of the magnetic structures for 
CoV$_{2}$O$_{4}$.
}
\end{figure}

To investigate how the magnetic state evolves with the changes in the crystal structure, neutron diffraction experiments were performed on polycrystalline and single crystal samples of CoV$_2$O$_4$. 
Figure 3 (a) 
shows the neutron powder diffraction data collected at 5 K ($<T_{\rm C}$) and 180 K ($>T_{\rm C}$). 
The overall crystal 
structure remains cubic with $Fd\overline{3}m$ symmetry down to 5 K. An obvious difference between the 5 K and 180 K data is the strong (111) Bragg intensity at 5 K, which is due to the  ferrimagnetic order with the characteristic wave vector of ${\bf k}_m=(0,0,0)$. The refinement of the diffraction data at 5 K indicate that the Co$^{2+}$ magnetic moments 
 are ferromagnetically aligned along one principal axis of the cubic spinel (whose 
 direction is defined as the $c$ axis) and the c-component of the V$^{3+}$ moments are antiparallel to the Co$^{2+}$ moments. In addition, the V moments are canted from the $c$ axis by $\sim$ 20(2)$^{\circ}$ to 
 the $\left< 110 \right>$ direction, and the $ab$-plane components of the neighboring V$^{3+}$ moments are antiferromagnetically aligned with each other. This magnetic structure is reproducible by the ${\Gamma}_9$ irreducible 
 representation for the $Fd\overline{3}m$ space group with ${\bf k}_m=(0,0,0)$, as illustrated in Figs. 3 (b). This magnetic structure 
 is similar to that of FeV$_{2}$O$_{4}$ at low temperatures \cite{MacDougall12}. The magnitude of the ordered Co moment at 5 K, $\left<M_{\rm Co}\right>=2.89(3)\mu_{\rm B}$, is close to the expected value for the high-spin state of 
Co$^{2+}$ (3 $\mu_{\rm B}$), while that of the V moment 
$\left<M_{\rm V}\right>= 0.71(3)\mu_{\rm B}$, is much less than the expected value for V$^{3+}$ (2 $\mu_{\rm B}$) when it is fully polarized. The  
reduction of the V moment is due to strong frustration in the pyrochlore lattice of V ions, which is common in vanadium spinels.  
\cite{lee04,MacDougall12,Garlea08}.

Four-circle neutron diffraction measurements on a single crystal (\#2) using a neutron wavelength of $\lambda=2.4$ \AA\ were also performed as a function of $T$ at several different Bragg {\bf Q} points. As shown in Fig. 4 (a) and (b), upon cooling, most of the Bragg peaks such as (400), (202), (313), (511) and (111) increase below 169 K $\sim$ $T_{\rm C}$, exhibit a broad maximum at $\sim$ 100 K, and a dip at $\sim$ 40 K. This $T$-dependence coincides with the $T$-dependence of the strain; $\Delta L_{\parallel}/L (H=0.2$ T) exhibits a broad peak at $\sim$ 100 K, and below $\sim 40$ K $\Delta L_{\perp}/L (H=0.2$ T) becomes larger than $\Delta L_{\parallel}/L (H=0.2$ T). An exception is the (002) peak that is a forbidden nuclear peak by the $Fd\overline{3}m$ symmetry, and thus expected to be purely magnetic. The (002) peak exhibits a gradual increase below $T_{\rm C}$ down to 10 K with a very weak dip at $\sim 40$ K.

\begin{figure}
\includegraphics[scale=0.43]{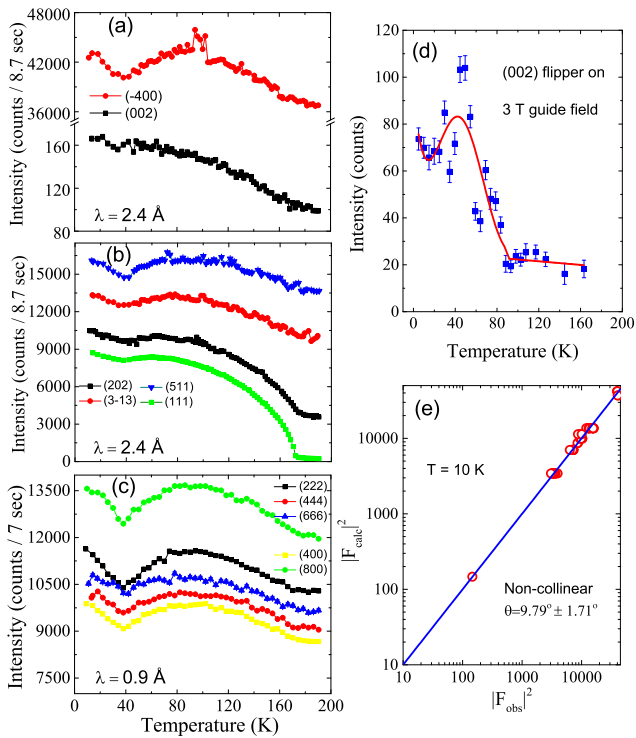}
\caption{(Color online) 
(a)(b) Temperature dependence of the measured intensity of main magnetic Bragg peaks 
of CoV$_2$O$_4$ single crystal (\#2) measured using neutron wavelength 
$\lambda$ = 2.4 \AA (c) Temperature dependence of the measured intensity of selected 
Bragg peaks with large absolute value of momentum transfer $Q$ measured using 
neutron wavelength $\lambda$ =0.9 \AA\ (d) Temperature 
dependence of the spin-flip intensity of (002) magnetic Bragg peak measured using 
polarized neutron with a vertical guide field of 3 T. Red line is a guide to the eye. (e) Single Crystal refinement of 
CoV$_2$O$_4$ (\#2) at 10 K : Measured ($y$ axis) and calculated ($x$ axis) values for the 
absolute nuclear and magnetic structure factors of the Bragg peaks.  
}
\end{figure}

In order to identify the $T$-dependences of the nuclear and magnetic contributions for the Bragg peaks, we have performed two different measurements. Firstly, unpolarized four-circle diffraction using a neutron wavelength of $\lambda=0.9$ \AA\ was performed to reach high {\bf Q} values where magnetic contributions are negligible due to the fall-off of the magnetic form factors of the Co$^{2+}$ and V$^{3+}$ ions. As shown in Fig. 4 (c), the high $Q$ Bragg peaks studied also exhibit similar $T$-dependences, and the dip at $\sim$ 40 K is more pronounced than the low $Q$ Bragg peaks shown in Fig. 4 (a) and (b). Thus, the observed dip of the Bragg peak intensities at $T_{p}$ is not entirely caused by the 
change in the magnetic structure. A possible origin is the change in the extinction effect 
on the diffraction intensities caused by a change in the mosaic structures \cite{note2}. In 
CoV$_{2}$O$_{4}$, such an extinction effect is the largest at $T_{p}$ at which the strain, $\Delta L/L$, is minimal, i.e., the distortion is minimal.

Secondly, we performed three-axis polarized neutron scattering measurements for the (002) Bragg peak. The crystal was aligned in the ($HHL$) scattering plane and a vertical guide field of 3 T was applied along the [1,-1,0] direction. The spin-flip scattering of the (002) Bragg reflection, shown in Fig. 4 (d) is proportional to the square of the antiferromagnetic in-plane $(110)$ moments of the V$^{3+}$ ions. Fig. 4 (d) shows the data as a function of $T$. The (002) intensity remains almost zero for $90$ K $< T < T_{\rm C}$, and below $\sim$ 90 K it increases sharply and exhibits a maximum at $\sim 40$ K. The unpolarized and polarized neutron scattering data indicate that upon cooling below $T_{\rm C}$ the V$^{3+}$ and Co$^{2+}$ moments order ferrimagnetically and upon further cooling the V$^{3+}$ moments start canting at $T \sim 90$ K. By refining the 10 K single crystal diffraction data (see Fig. 4 (e)), the canting angle of the V$^{3+}$ moments is determined to be $\sim$ 10(2)$^{\circ}$, which is close to the value of 20(2)$^{\circ}$ obtained from the powder diffraction. 
Similar cantings of the V$^{3+}$ moments have been observed in MnV$_{2}$O$_{4}$ with the canting angle of $\sim 65^{\circ}$\cite{Chung08,Garlea08} and FeV$_{2}$O$_{4}$ with that of $\sim 55^{\circ}$\cite{MacDougall12}, and they were suggested to closely relate to an orbital order of V$^{3+}$ ($t_{2g}^2$) ions and the resulting change in their magnetic interactions. The fact that in CoV$_{2}$O$_{4}$ the canting occurs at temperatures where the strain $\Delta L/L$ is maximized ($T_{\rm max} \sim 90$ K for the \#2 crystal 
\cite{note0}) indicates that an orbital order appears at $\sim$ 100 K as well. Unlike in the insulating compounds, MnV$_{2}$O$_{4}$ and FeV$_{2}$O$_{4}$, where the orbital order occurs in a first order fashion, however, the structural 
distortion of CoV$_2$O$_4$ is subtle and gradual as a function of temperature. Thus, we conclude that in CoV$_2$O$_4$ its close proximity to the itineracy incompletely suppresses the orbital degree of freedom of the V$^{3+}$ ions, leading to an orbital glass state. 

In summary, our bulk magnetization, magnetostriction, and neutron scattering data obtained from polycrystalline and single crystals of the nearly metallic vanadium spinel, CoV$_2$O$_4$, show that upon cooling the system undergoes two successive second order phase transitions at $T_{\rm C} \sim 160$ K from a paramagnet to a collinear ferrimagnet, and at $\sim$ 100 K to a noncollinear ferrimagnet and orbital glassy state. Our results suggest that the combination of the magnetostriction and polarized and unpolarized neutron scattering techniques may be powerful in studying the subtle interplays between the orbital and spin degrees of freedom in other materials as well in which fluctuations of the V orbitals are supposed to play an important role.

The work at Waseda university was partly supported by JSPS 
KAKENHI Grant No. 25287090. Research at UVA was supported by the US 
Department of Energy, Office of Basic Energy Sciences,Division of Materials Sciences 
and Engineering, under Award No. DE-FG02-07ER46384. We acknowledge the support of the National Institute of Standards and Technology, U. S. 
Department of Commerce, in providing the neutron research facilities for powder 
diffraction measurements used in this work. This research at ORNL's High Flux Isotope Reactor was sponsored by the Scientific User Facilities Division, Office of Basic Energy Sciences, US Department of Energy.


\begin{thebibliography}{30}
\expandafter\ifx\csname natexlab\endcsname\relax\def\natexlab#1{#1}\fi
\expandafter\ifx\csname bibnamefont\endcsname\relax
  \def\bibnamefont#1{#1}\fi
\expandafter\ifx\csname bibfnamefont\endcsname\relax
  \def\bibfnamefont#1{#1}\fi
\expandafter\ifx\csname citenamefont\endcsname\relax
  \def\citenamefont#1{#1}\fi
\expandafter\ifx\csname url\endcsname\relax
  \def\url#1{\texttt{#1}}\fi
\expandafter\ifx\csname urlprefix\endcsname\relax\def\urlprefix{URL }\fi
\providecommand{\bibinfo}[2]{#2}
\providecommand{\eprint}[2][]{\url{#2}}

\bibitem[{\citenamefont{Ueda et~al.}(1997)\citenamefont{Ueda, Fujiwara, and
  Yasuoka}}]{ueda97}
\bibinfo{author}{\bibfnamefont{Y.}~\bibnamefont{Ueda}},
  \bibinfo{author}{\bibfnamefont{N.}~\bibnamefont{Fujiwara}}, \bibnamefont{and}
  \bibinfo{author}{\bibfnamefont{H.}~\bibnamefont{Yasuoka}},
  \bibinfo{journal}{J. Phys. Soc. Jpn.} \textbf{\bibinfo{volume}{66}},
  \bibinfo{pages}{778} (\bibinfo{year}{1997}).

\bibitem[{\citenamefont{Lee et~al.}(2004)}]{lee04}
\bibinfo{author}{\bibfnamefont{S.-H.} \bibnamefont{Lee}} \bibnamefont{et~al.},
  \bibinfo{journal}{Phys. Rev. Lett.} \textbf{\bibinfo{volume}{93}},
  \bibinfo{pages}{156407} (\bibinfo{year}{2004}).

\bibitem[{\citenamefont{Zhang et~al.}(2006)}]{zhang06}
\bibinfo{author}{\bibfnamefont{Z.}~\bibnamefont{Zhang}} \bibnamefont{et~al.},
  \bibinfo{journal}{Phys. Rev. B} \textbf{\bibinfo{volume}{74}},
  \bibinfo{pages}{014108} (\bibinfo{year}{2006}).

\bibitem[{\citenamefont{Suzuki et~al.}(2007)\citenamefont{Suzuki, Katsumura,
  Taniguchi, Arima, and Katsufuji}}]{Suzuki07}
\bibinfo{author}{\bibfnamefont{T.}~\bibnamefont{Suzuki}},
  \bibinfo{author}{\bibfnamefont{M.}~\bibnamefont{Katsumura}},
  \bibinfo{author}{\bibfnamefont{K.}~\bibnamefont{Taniguchi}},
  \bibinfo{author}{\bibfnamefont{T.}~\bibnamefont{Arima}}, \bibnamefont{and}
  \bibinfo{author}{\bibfnamefont{T.}~\bibnamefont{Katsufuji}},
  \bibinfo{journal}{Phys. Rev. Lett.} \textbf{\bibinfo{volume}{98}},
  \bibinfo{pages}{127203} (\bibinfo{year}{2007}).

\bibitem[{\citenamefont{Zhou et~al.}(2007)\citenamefont{Zhou, Lu, and
  Wiebe}}]{Zhou07}
\bibinfo{author}{\bibfnamefont{H.~D.} \bibnamefont{Zhou}},
  \bibinfo{author}{\bibfnamefont{J.}~\bibnamefont{Lu}}, \bibnamefont{and}
  \bibinfo{author}{\bibfnamefont{C.~R.} \bibnamefont{Wiebe}},
  \bibinfo{journal}{Phys. Rev. B} \textbf{\bibinfo{volume}{76}},
  \bibinfo{pages}{174403} (\bibinfo{year}{2007}).

\bibitem[{\citenamefont{Chung et~al.}(2008)\citenamefont{Chung, Kim, Lee, Sato,
  Suzuki, Katsumura, and Katsufuji}}]{Chung08}
\bibinfo{author}{\bibfnamefont{J.~H.} \bibnamefont{Chung}},
  \bibinfo{author}{\bibfnamefont{J.~H.} \bibnamefont{Kim}},
  \bibinfo{author}{\bibfnamefont{S.~H.} \bibnamefont{Lee}},
  \bibinfo{author}{\bibfnamefont{T.~J.} \bibnamefont{Sato}},
  \bibinfo{author}{\bibfnamefont{T.}~\bibnamefont{Suzuki}},
  \bibinfo{author}{\bibfnamefont{M.}~\bibnamefont{Katsumura}},
  \bibnamefont{and}
  \bibinfo{author}{\bibfnamefont{T.}~\bibnamefont{Katsufuji}},
  \bibinfo{journal}{Phys. Rev. B} \textbf{\bibinfo{volume}{77}},
  \bibinfo{pages}{054412} (\bibinfo{year}{2008}).

\bibitem[{\citenamefont{Garlea et~al.}(2008)\citenamefont{Garlea, Jin, Mandrus,
  Roessli, Huang, Miller, Schultz, and Nagler}}]{Garlea08}
\bibinfo{author}{\bibfnamefont{V.~O.} \bibnamefont{Garlea}},
  \bibinfo{author}{\bibfnamefont{R.}~\bibnamefont{Jin}},
  \bibinfo{author}{\bibfnamefont{D.}~\bibnamefont{Mandrus}},
  \bibinfo{author}{\bibfnamefont{B.}~\bibnamefont{Roessli}},
  \bibinfo{author}{\bibfnamefont{Q.}~\bibnamefont{Huang}},
  \bibinfo{author}{\bibfnamefont{M.}~\bibnamefont{Miller}},
  \bibinfo{author}{\bibfnamefont{A.~J.} \bibnamefont{Schultz}},
  \bibnamefont{and} \bibinfo{author}{\bibfnamefont{S.~E.}
  \bibnamefont{Nagler}}, \bibinfo{journal}{Phys. Rev. Lett.}
  \textbf{\bibinfo{volume}{100}}, \bibinfo{pages}{066404}
  (\bibinfo{year}{2008}).

\bibitem[{\citenamefont{Hardy et~al.}(2008)\citenamefont{Hardy, Br{\'e}ard, and
  Martin}}]{Hardy08}
\bibinfo{author}{\bibfnamefont{V.}~\bibnamefont{Hardy}},
  \bibinfo{author}{\bibfnamefont{Y.}~\bibnamefont{Br{\'e}ard}},
  \bibnamefont{and} \bibinfo{author}{\bibfnamefont{C.}~\bibnamefont{Martin}},
  \bibinfo{journal}{Phys. Rev. B} \textbf{\bibinfo{volume}{78}},
  \bibinfo{pages}{024406} (\bibinfo{year}{2008}).

\bibitem[{\citenamefont{Baek et~al.}(2009)\citenamefont{Baek, Curro, Choi,
  Reyes, Kuhns, Zhou, and Wiebe}}]{Baek09}
\bibinfo{author}{\bibfnamefont{S.-H.} \bibnamefont{Baek}},
  \bibinfo{author}{\bibfnamefont{N.~J.} \bibnamefont{Curro}},
  \bibinfo{author}{\bibfnamefont{K.-Y.} \bibnamefont{Choi}},
  \bibinfo{author}{\bibfnamefont{A.~P.} \bibnamefont{Reyes}},
  \bibinfo{author}{\bibfnamefont{P.~L.} \bibnamefont{Kuhns}},
  \bibinfo{author}{\bibfnamefont{H.~D.} \bibnamefont{Zhou}}, \bibnamefont{and}
  \bibinfo{author}{\bibfnamefont{C.~R.} \bibnamefont{Wiebe}},
  \bibinfo{journal}{Phys. Rev. B} \textbf{\bibinfo{volume}{80}},
  \bibinfo{pages}{140406} (\bibinfo{year}{2009}).

\bibitem[{\citenamefont{Sarkar et~al.}(2009)\citenamefont{Sarkar, Maitra,
  Valent\'{i}, and Saha-Dasgupta}}]{Sarkar09}
\bibinfo{author}{\bibfnamefont{S.}~\bibnamefont{Sarkar}},
  \bibinfo{author}{\bibfnamefont{T.}~\bibnamefont{Maitra}},
  \bibinfo{author}{\bibfnamefont{R.}~\bibnamefont{Valent\'{i}}},
  \bibnamefont{and}
  \bibinfo{author}{\bibfnamefont{T.}~\bibnamefont{Saha-Dasgupta}},
  \bibinfo{journal}{Phys. Rev. Lett.} \textbf{\bibinfo{volume}{102}},
  \bibinfo{pages}{216405} (\bibinfo{year}{2009}).

\bibitem[{\citenamefont{Chern et~al.}(2010)\citenamefont{Chern, Perkins, and
  Hao}}]{Chern10}
\bibinfo{author}{\bibfnamefont{G.-W.} \bibnamefont{Chern}},
  \bibinfo{author}{\bibfnamefont{N.}~\bibnamefont{Perkins}}, \bibnamefont{and}
  \bibinfo{author}{\bibfnamefont{Z.}~\bibnamefont{Hao}},
  \bibinfo{journal}{Phys. Rev. B} \textbf{\bibinfo{volume}{81}},
  \bibinfo{pages}{125127} (\bibinfo{year}{2010}).

\bibitem[{\citenamefont{Nii et~al.}(2013)\citenamefont{Nii, Abe, and
  Arima}}]{Nii13}
\bibinfo{author}{\bibfnamefont{Y.}~\bibnamefont{Nii}},
  \bibinfo{author}{\bibfnamefont{N.}~\bibnamefont{Abe}}, \bibnamefont{and}
  \bibinfo{author}{\bibfnamefont{T.}~\bibnamefont{Arima}},
  \bibinfo{journal}{Phys. Rev. B} \textbf{\bibinfo{volume}{87}},
  \bibinfo{pages}{085111} (\bibinfo{year}{2013}).

\bibitem[{\citenamefont{Gleason et~al.}(2014)\citenamefont{Gleason, Byrum, Gim,
  Thaler, Abbamonte, MacDougall, Martin, Zhou, and Cooper}}]{Gleason14}
\bibinfo{author}{\bibfnamefont{S.~L.} \bibnamefont{Gleason}},
  \bibinfo{author}{\bibfnamefont{T.}~\bibnamefont{Byrum}},
  \bibinfo{author}{\bibfnamefont{Y.}~\bibnamefont{Gim}},
  \bibinfo{author}{\bibfnamefont{A.}~\bibnamefont{Thaler}},
  \bibinfo{author}{\bibfnamefont{P.}~\bibnamefont{Abbamonte}},
  \bibinfo{author}{\bibfnamefont{G.~J.} \bibnamefont{MacDougall}},
  \bibinfo{author}{\bibfnamefont{L.~W.} \bibnamefont{Martin}},
  \bibinfo{author}{\bibfnamefont{H.~D.} \bibnamefont{Zhou}}, \bibnamefont{and}
  \bibinfo{author}{\bibfnamefont{S.~L.} \bibnamefont{Cooper}},
  \bibinfo{journal}{Phys. Rev. B} \textbf{\bibinfo{volume}{89}},
  \bibinfo{pages}{134402} (\bibinfo{year}{2014}).

\bibitem[{\citenamefont{Katsufuji et~al.}(2008)\citenamefont{Katsufuji, Suzuki,
  Takei, Shingu, Kato, Osaka, Takata, Sagayama, and Arima}}]{Katsufuji08}
\bibinfo{author}{\bibfnamefont{T.}~\bibnamefont{Katsufuji}},
  \bibinfo{author}{\bibfnamefont{T.}~\bibnamefont{Suzuki}},
  \bibinfo{author}{\bibfnamefont{H.}~\bibnamefont{Takei}},
  \bibinfo{author}{\bibfnamefont{M.}~\bibnamefont{Shingu}},
  \bibinfo{author}{\bibfnamefont{K.}~\bibnamefont{Kato}},
  \bibinfo{author}{\bibfnamefont{K.}~\bibnamefont{Osaka}},
  \bibinfo{author}{\bibfnamefont{M.}~\bibnamefont{Takata}},
  \bibinfo{author}{\bibfnamefont{H.}~\bibnamefont{Sagayama}}, \bibnamefont{and}
  \bibinfo{author}{\bibfnamefont{T.}~\bibnamefont{Arima}}, \bibinfo{journal}{J.
  Phys. Soc. Jpn.} \textbf{\bibinfo{volume}{77}}, \bibinfo{pages}{053708}
  (\bibinfo{year}{2008}).

\bibitem[{\citenamefont{Sarkar and Saha-Dasgupta}(2011)}]{Sarkar11}
\bibinfo{author}{\bibfnamefont{S.}~\bibnamefont{Sarkar}} \bibnamefont{and}
  \bibinfo{author}{\bibfnamefont{T.}~\bibnamefont{Saha-Dasgupta}},
  \bibinfo{journal}{Phys. Rev. B} \textbf{\bibinfo{volume}{84}},
  \bibinfo{pages}{235112} (\bibinfo{year}{2011}).

\bibitem[{\citenamefont{MacDougall et~al.}(2012)\citenamefont{MacDougall,
  Garlea, Aczel, Zhou, and Nagler}}]{MacDougall12}
\bibinfo{author}{\bibfnamefont{G.~J.} \bibnamefont{MacDougall}},
  \bibinfo{author}{\bibfnamefont{V.~O.} \bibnamefont{Garlea}},
  \bibinfo{author}{\bibfnamefont{A.~A.} \bibnamefont{Aczel}},
  \bibinfo{author}{\bibfnamefont{H.~D.} \bibnamefont{Zhou}}, \bibnamefont{and}
  \bibinfo{author}{\bibfnamefont{S.~E.} \bibnamefont{Nagler}},
  \bibinfo{journal}{Phys. Rev. B} \textbf{\bibinfo{volume}{86}},
  \bibinfo{pages}{060414} (\bibinfo{year}{2012}).

\bibitem[{\citenamefont{Zhang et~al.}(2012)\citenamefont{Zhang, Singh, Guillou,
  Simon, Breard, Caignaert, and Hardy}}]{Zhang12}
\bibinfo{author}{\bibfnamefont{Q.}~\bibnamefont{Zhang}},
  \bibinfo{author}{\bibfnamefont{K.}~\bibnamefont{Singh}},
  \bibinfo{author}{\bibfnamefont{F.}~\bibnamefont{Guillou}},
  \bibinfo{author}{\bibfnamefont{C.}~\bibnamefont{Simon}},
  \bibinfo{author}{\bibfnamefont{Y.}~\bibnamefont{Breard}},
  \bibinfo{author}{\bibfnamefont{V.}~\bibnamefont{Caignaert}},
  \bibnamefont{and} \bibinfo{author}{\bibfnamefont{V.}~\bibnamefont{Hardy}},
  \bibinfo{journal}{Phys. Rev. B} \textbf{\bibinfo{volume}{85}},
  \bibinfo{pages}{054405} (\bibinfo{year}{2012}).

\bibitem[{\citenamefont{Nii et~al.}(2012)\citenamefont{Nii, Sagayama, Arima,
  Aoyagi, Sakai, Maki, Nishibori, Sawa, Sugimoto, Ohsumi et~al.}}]{Nii12}
\bibinfo{author}{\bibfnamefont{Y.}~\bibnamefont{Nii}},
  \bibinfo{author}{\bibfnamefont{H.}~\bibnamefont{Sagayama}},
  \bibinfo{author}{\bibfnamefont{T.}~\bibnamefont{Arima}},
  \bibinfo{author}{\bibfnamefont{S.}~\bibnamefont{Aoyagi}},
  \bibinfo{author}{\bibfnamefont{R.}~\bibnamefont{Sakai}},
  \bibinfo{author}{\bibfnamefont{S.}~\bibnamefont{Maki}},
  \bibinfo{author}{\bibfnamefont{E.}~\bibnamefont{Nishibori}},
  \bibinfo{author}{\bibfnamefont{H.}~\bibnamefont{Sawa}},
  \bibinfo{author}{\bibfnamefont{K.}~\bibnamefont{Sugimoto}},
  \bibinfo{author}{\bibfnamefont{H.}~\bibnamefont{Ohsumi}},
  \bibnamefont{et~al.}, \bibinfo{journal}{Phys. Rev. B}
  \textbf{\bibinfo{volume}{86}}, \bibinfo{pages}{125142}
  (\bibinfo{year}{2012}).

\bibitem[{\citenamefont{Kang et~al.}(2012)\citenamefont{Kang, Hwang, Kim, Lee,
  Kim, Kim, Kwon, Lee, Kim, Ueno et~al.}}]{Kang12}
\bibinfo{author}{\bibfnamefont{J.-S.} \bibnamefont{Kang}},
  \bibinfo{author}{\bibfnamefont{J.}~\bibnamefont{Hwang}},
  \bibinfo{author}{\bibfnamefont{D.~H.} \bibnamefont{Kim}},
  \bibinfo{author}{\bibfnamefont{E.}~\bibnamefont{Lee}},
  \bibinfo{author}{\bibfnamefont{W.~C.} \bibnamefont{Kim}},
  \bibinfo{author}{\bibfnamefont{C.~S.} \bibnamefont{Kim}},
  \bibinfo{author}{\bibfnamefont{S.}~\bibnamefont{Kwon}},
  \bibinfo{author}{\bibfnamefont{S.}~\bibnamefont{Lee}},
  \bibinfo{author}{\bibfnamefont{J.-Y.} \bibnamefont{Kim}},
  \bibinfo{author}{\bibfnamefont{T.}~\bibnamefont{Ueno}}, \bibnamefont{et~al.},
  \bibinfo{journal}{Phys. Rev. B} \textbf{\bibinfo{volume}{85}},
  \bibinfo{pages}{165136} (\bibinfo{year}{2012}).

\bibitem[{\citenamefont{Kawaguchi et~al.}(2013)\citenamefont{Kawaguchi,
  Ishibashi, Nishihara, Miyagawa, Inoue, Mori, and Kubota}}]{Kawaguchi13}
\bibinfo{author}{\bibfnamefont{S.}~\bibnamefont{Kawaguchi}},
  \bibinfo{author}{\bibfnamefont{H.}~\bibnamefont{Ishibashi}},
  \bibinfo{author}{\bibfnamefont{S.}~\bibnamefont{Nishihara}},
  \bibinfo{author}{\bibfnamefont{M.}~\bibnamefont{Miyagawa}},
  \bibinfo{author}{\bibfnamefont{K.}~\bibnamefont{Inoue}},
  \bibinfo{author}{\bibfnamefont{S.}~\bibnamefont{Mori}}, \bibnamefont{and}
  \bibinfo{author}{\bibfnamefont{Y.}~\bibnamefont{Kubota}},
  \bibinfo{journal}{J. Phys.: Condens. Matter} \textbf{\bibinfo{volume}{25}},
  \bibinfo{pages}{416005} (\bibinfo{year}{2013}).

\bibitem[{\citenamefont{Huang et~al.}(2014)\citenamefont{Huang, Luo, Hu, Tan,
  Liu, Yuan, Chen, Song, and Sun}}]{Huang14}
\bibinfo{author}{\bibfnamefont{Z.~H.} \bibnamefont{Huang}},
  \bibinfo{author}{\bibfnamefont{X.}~\bibnamefont{Luo}},
  \bibinfo{author}{\bibfnamefont{L.}~\bibnamefont{Hu}},
  \bibinfo{author}{\bibfnamefont{S.~G.} \bibnamefont{Tan}},
  \bibinfo{author}{\bibfnamefont{Y.}~\bibnamefont{Liu}},
  \bibinfo{author}{\bibfnamefont{B.}~\bibnamefont{Yuan}},
  \bibinfo{author}{\bibfnamefont{J.}~\bibnamefont{Chen}},
  \bibinfo{author}{\bibfnamefont{W.~H.} \bibnamefont{Song}}, \bibnamefont{and}
  \bibinfo{author}{\bibfnamefont{Y.~P.} \bibnamefont{Sun}},
  \bibinfo{journal}{J. Appl. Phys.} \textbf{\bibinfo{volume}{115}},
  \bibinfo{pages}{034903} (\bibinfo{year}{2014}).

\bibitem[{\citenamefont{Choudhury et~al.}(2014)\citenamefont{Choudhury, Suzuki,
  Okuyama, Morikawa, Kato, Takata, Kobayashi, Kumai, Nakao, Murakami
  et~al.}}]{Choudhury14}
\bibinfo{author}{\bibfnamefont{D.}~\bibnamefont{Choudhury}},
  \bibinfo{author}{\bibfnamefont{T.}~\bibnamefont{Suzuki}},
  \bibinfo{author}{\bibfnamefont{D.}~\bibnamefont{Okuyama}},
  \bibinfo{author}{\bibfnamefont{D.}~\bibnamefont{Morikawa}},
  \bibinfo{author}{\bibfnamefont{K.}~\bibnamefont{Kato}},
  \bibinfo{author}{\bibfnamefont{M.}~\bibnamefont{Takata}},
  \bibinfo{author}{\bibfnamefont{K.}~\bibnamefont{Kobayashi}},
  \bibinfo{author}{\bibfnamefont{R.}~\bibnamefont{Kumai}},
  \bibinfo{author}{\bibfnamefont{H.}~\bibnamefont{Nakao}},
  \bibinfo{author}{\bibfnamefont{Y.}~\bibnamefont{Murakami}},
  \bibnamefont{et~al.}, \bibinfo{journal}{Phys. Rev. B}
  \textbf{\bibinfo{volume}{89}}, \bibinfo{pages}{104427}
  (\bibinfo{year}{2014}).

\bibitem[{\citenamefont{Kismarahardja et~al.}(2011)\citenamefont{Kismarahardja,
  Brooks, Kiswandhi, Matsubayashi, Yamanaka, Uwatoko, Whalen, Siegrist, and
  Zhou}}]{Kismarahardja11}
\bibinfo{author}{\bibfnamefont{A.}~\bibnamefont{Kismarahardja}},
  \bibinfo{author}{\bibfnamefont{J.~S.} \bibnamefont{Brooks}},
  \bibinfo{author}{\bibfnamefont{A.}~\bibnamefont{Kiswandhi}},
  \bibinfo{author}{\bibfnamefont{K.}~\bibnamefont{Matsubayashi}},
  \bibinfo{author}{\bibfnamefont{R.}~\bibnamefont{Yamanaka}},
  \bibinfo{author}{\bibfnamefont{Y.}~\bibnamefont{Uwatoko}},
  \bibinfo{author}{\bibfnamefont{J.}~\bibnamefont{Whalen}},
  \bibinfo{author}{\bibfnamefont{T.}~\bibnamefont{Siegrist}}, \bibnamefont{and}
  \bibinfo{author}{\bibfnamefont{H.~D.} \bibnamefont{Zhou}},
  \bibinfo{journal}{Phys. Rev. Lett.} \textbf{\bibinfo{volume}{106}},
  \bibinfo{pages}{056602} (\bibinfo{year}{2011}).

\bibitem[{\citenamefont{Kiswandhi et~al.}(2011)\citenamefont{Kiswandhi, Brooks,
  Lu, Whalen, Siegrist, and Zhou}}]{Kiswandhi11}
\bibinfo{author}{\bibfnamefont{A.}~\bibnamefont{Kiswandhi}},
  \bibinfo{author}{\bibfnamefont{J.~S.} \bibnamefont{Brooks}},
  \bibinfo{author}{\bibfnamefont{J.}~\bibnamefont{Lu}},
  \bibinfo{author}{\bibfnamefont{J.}~\bibnamefont{Whalen}},
  \bibinfo{author}{\bibfnamefont{T.}~\bibnamefont{Siegrist}}, \bibnamefont{and}
  \bibinfo{author}{\bibfnamefont{H.~D.} \bibnamefont{Zhou}},
  \bibinfo{journal}{Phys. Rev. B} \textbf{\bibinfo{volume}{84}},
  \bibinfo{pages}{205138} (\bibinfo{year}{2011}).

\bibitem[{\citenamefont{Huang et~al.}(2012)\citenamefont{Huang, Yang, and
  Zhang}}]{Huang12}
\bibinfo{author}{\bibfnamefont{Y.}~\bibnamefont{Huang}},
  \bibinfo{author}{\bibfnamefont{Z.}~\bibnamefont{Yang}}, \bibnamefont{and}
  \bibinfo{author}{\bibfnamefont{Y.}~\bibnamefont{Zhang}}, \bibinfo{journal}{J.
  Phys., Condens. Matter} \textbf{\bibinfo{volume}{24}},
  \bibinfo{pages}{056003} (\bibinfo{year}{2012}).

\bibitem[{\citenamefont{Kaur et~al.}(2014)\citenamefont{Kaur, Maitra, and
  Nautiyal}}]{Kaur14}
\bibinfo{author}{\bibfnamefont{R.}~\bibnamefont{Kaur}},
  \bibinfo{author}{\bibfnamefont{T.}~\bibnamefont{Maitra}}, \bibnamefont{and}
  \bibinfo{author}{\bibfnamefont{T.}~\bibnamefont{Nautiyal}},
  \bibinfo{journal}{J. Phys.: Condens. Matter} \textbf{\bibinfo{volume}{26}},
  \bibinfo{pages}{045505} (\bibinfo{year}{2014}).

\bibitem[{\citenamefont{Kismarahardja et~al.}(2013)\citenamefont{Kismarahardja,
  Brooks, Zhou, Choi, Matsubayashi, , and Uwatoko}}]{Kismarahardja13}
\bibinfo{author}{\bibfnamefont{A.}~\bibnamefont{Kismarahardja}},
  \bibinfo{author}{\bibfnamefont{J.~S.} \bibnamefont{Brooks}},
  \bibinfo{author}{\bibfnamefont{H.~D.} \bibnamefont{Zhou}},
  \bibinfo{author}{\bibfnamefont{E.~S.} \bibnamefont{Choi}},
  \bibinfo{author}{\bibfnamefont{K.}~\bibnamefont{Matsubayashi}}, ,
  \bibnamefont{and} \bibinfo{author}{\bibfnamefont{Y.}~\bibnamefont{Uwatoko}},
  \bibinfo{journal}{Phys. Rev. B} \textbf{\bibinfo{volume}{87}},
  \bibinfo{pages}{054432} (\bibinfo{year}{2013}).

\bibitem[{\citenamefont{Ma et~al.}(2015)\citenamefont{Ma, Lee, Hahn, Hong, Cao,
  Aczel, Dun, Stone, Tian, Qiu et~al.}}]{Ma14}
\bibinfo{author}{\bibfnamefont{J.}~\bibnamefont{Ma}},
  \bibinfo{author}{\bibfnamefont{J.~H.} \bibnamefont{Lee}},
  \bibinfo{author}{\bibfnamefont{S.~E.} \bibnamefont{Hahn}},
  \bibinfo{author}{\bibfnamefont{T.}~\bibnamefont{Hong}},
  \bibinfo{author}{\bibfnamefont{H.~B.} \bibnamefont{Cao}},
  \bibinfo{author}{\bibfnamefont{A.~A.} \bibnamefont{Aczel}},
  \bibinfo{author}{\bibfnamefont{Z.~L.} \bibnamefont{Dun}},
  \bibinfo{author}{\bibfnamefont{M.~B.} \bibnamefont{Stone}},
  \bibinfo{author}{\bibfnamefont{W.}~\bibnamefont{Tian}},
  \bibinfo{author}{\bibfnamefont{Y.}~\bibnamefont{Qiu}}, \bibnamefont{et~al.},
  \bibinfo{journal}{Phys. Rev. B} \textbf{\bibinfo{volume}{91}},
  \bibinfo{pages}{020407} (\bibinfo{year}{2015}).

\bibitem[{not({\natexlab{a}})}]{note0}
\bibinfo{note}{$T$ dependence of $M$ and $\Delta L/L$ for other single crystals
  and polycrystalline Co$_{1+x}$V$_{2-x}$O$_{4}$ are shown and discussed in the
  supplementary material.}

\bibitem[{not({\natexlab{b}})}]{note2}
\bibinfo{note}{In neuron scattering of a real crystal, which can be regarded as
  a bunch of many crystal blocks with {\it nearly} the same orientation, an
  incident neutron beam is reflected from many crystal blocks if there is small
  misorientation between the blocks. However, if the orientation of the crystal
  blocks are perfect, an upstream crystal block depletes the beam before it
  reaches a downstream crystal block that has exactly the same orientation and
  is supposed to reflect the same beam as the upstream crystal. This effect
  (extinction effect) reduces the scattering intensity of the beam.}

\end{thebibliography}
\end{document}